\documentclass[doublecol,figures]{epl2}
\usepackage{bm}
\usepackage{amsmath,amssymb}
\usepackage{float}
\usepackage{color}
\usepackage{latexsym}
\usepackage{setspace}
\usepackage{here}
\usepackage{caption}
\usepackage{hyperref}
\hypersetup{
     colorlinks   = true,
     linkcolor    = black,
     citecolor    = black,
     urlcolor = blue
}

\title{Emergent isotropy of a wave-turbulent cascade in the Gross-Pitaevskii model}

\author{Yuto Sano\inst{1}  \and Nir Navon\inst{2,3} \and Makoto Tsubota\inst{4}}
\shortauthor{Yuto Sano \etal}

\institute{                    
  \inst{1} Department of Physics, Osaka City University, 3-3-138 Sugimoto,  Sumiyoshi-ku, Osaka 558-8585, Japan\\
  \inst{2} Department of Physics, Yale University, New Haven, Connecticut 06520, USA\\
  \inst{3} Yale Quantum Institute, Yale University, New Haven, Connecticut 06520, USA\\
  \inst{4} Department of Physics, Nambu Yoichiro Institute of Theoretical and Experimental Physics (NITEP), Osaka Metropolitan University, 3-3-138 Sugimoto, Sumiyoshi-ku, Osaka 558-8585, Japan\\
}


\abstract{
The restoration of symmetries is one of the most fascinating properties of turbulence. We report a study of the emergence of isotropy in the Gross-Pitaevskii model with anisotropic forcing. Inspired by recent experiments, we study the dynamics of a Bose-Einstein condensate in a cylindrical box driven along the symmetry axis of the trap by a spatially uniform force.
We introduce a measure of anisotropy $A(k,t)$ defined on the momentum distributions $n(\boldsymbol{k},t)$, and study the evolution of $A(k,t)$ and $n(\boldsymbol{k},t)$ as turbulence proceeds. As the system reaches a steady state, the anisotropy, large at low momenta because of the large-scale forcing, is greatly reduced at high momenta. While $n(\boldsymbol{k},t)$ exhibits a self-similar cascade front propagation, $A(k,t)$ decreases without such self-similar dynamics.
Finally, our numerical calculations show that the isotropy of the steady state is robust with respect to the amplitude of the drive.
}

\begin{document}

\maketitle

\section{Introduction}

Turbulence is an ubiquitous phenomenon in nonlinear science. Despite its complexity, turbulence is known to exhibit remarkably simple emergent features. One such feature is the statistical restoration of symmetries. Weak flows are typically sensitive to boundary conditions - even far from the boundaries -  and often break various symmetries (associated with the direction of the flow, for instance). On the other hand, at large fluid velocities, such broken symmetries are usually restored, in a statistical sense, at small length scales~\cite{frisch1995turbulence,davidson_turbulence}. The discovery of statistical restoration of symmetries and the emergence of universal laws form the backbone of our understanding of turbulence. The prime example is the observation of Kolmogorov's `$-5/3$' law \cite{obukhov1941energy} of homogeneous isotropic turbulence in anisotropically forced flows~\cite{frisch1995turbulence,davidson_turbulence}. 

The problem of `return to isotropy', i.e. how anisotropic forcing can lead to statistically isotropic turbulent fields, has been abundantly studied in hydrodynamic turbulence. For instance, quantities such as the Reynolds stress anisotropy tensor and the spectral anisotropy tensor of the energy spectrum have been introduced to investigate and classify turbulent flows~\cite{lumley1977return,yeung1991response,choi2001return,banerjee2007presentation}. Similar problems of `isotropization' of quantum fields have also been studied in the context of heavy-ion collisions~\cite{berges2005isotropization,berges2008bottom}.
    
The Gross-Pitaevskii (GP) model~\cite{tsatsos2016quantum,tsubota2017numerical} has been a popular tool to study turbulence, such as qualitative aspects of vortex-turbulent superfluids~\cite{nore1997kolmogorov,nore1997decaying,kobayashi2005kolmogorov2,kobayashi2005kolmogorov,parker2005emergence,kobayashi2007quantum,yepez2009superfluid}, turbulence in optical media \cite{dyachenko1992optical}, and wave turbulence in Bose-Einstein condensates (BEC)~\cite{lvov2003wave,zakharov2005dynamics,nazarenko2006wave,nazarenko2007freely,proment2009quantum}. The advent of ultracold gases as novel turbulent fluids~\cite{henn2009emergence,neely2013characteristics,seo2017observation,gauthier2019giant,johnstone2019evolution,navon2016emergence,navon2019synthetic} has rekindled the interest in the GP model~\cite{proment2012sustained,nazarenko2014bose,fujimoto2015bogoliubov,fujimoto2016direct,chantesana2019kinetic,mikheev2019low,semisalov2021numerical,zhu2021testing,griffin2021energy,shukla2022nonequilibrium}. While this model naturally describes the ground state and near-equilibrium properties of weakly interacting BECs, recent experiments have shown, surprisingly, that this model is also quantitatively useful in far-from-equilibrium regimes~\cite{johnstone2019evolution,navon2016emergence,navon2019synthetic,Gasenzer2017,Roati2021}. 
A key observation, in both experiments and GP simulations,  was the appearance of a statistically isotropic power-law momentum distribution under strongly anisotropic forcing~\cite{navon2016emergence}. 

Despite its popularity, little is known about the dynamic restoration of symmetries in the GP model. Inspired by the experiments of Refs.~\cite{navon2016emergence,navon2019synthetic}, we study here how isotropy emerges in a wave-turbulent cascade of the GP model. 

\section{Theoretical model}

Our theoretical model is the dimensionless GP equation
\begin{equation}
i\frac{\partial \psi(\bm r,\it t)}{\partial t}=\left (-{\nabla}^2+V(\bm r, t)+g |\psi(\bm r, t)|^2\right)\psi(\bm r,\it t) \,  
\label{eq:GP}
\end{equation}
describing the classical field $\psi({\bm r},\it t)$ of a weakly interacting Bose gas in a time-dependent external potential $V(\bm r,t)$, and $g$ is the dimensionless coupling constant. 

Following the experiments performed with $^{87}$Rb atoms trapped in optical boxes~\cite{navon2016emergence,navon2019synthetic}, we write the potential energy in the form
\begin{equation}
V({ \boldsymbol r}, t)=V_{\rm box} ({\boldsymbol r})+V_{\rm osc } (z, t ) + iV_{ \rm diss} ({\boldsymbol r}).
\label{eq:potential}
\end{equation}

\begin{figure}[h]
\includegraphics[width=8.0cm]{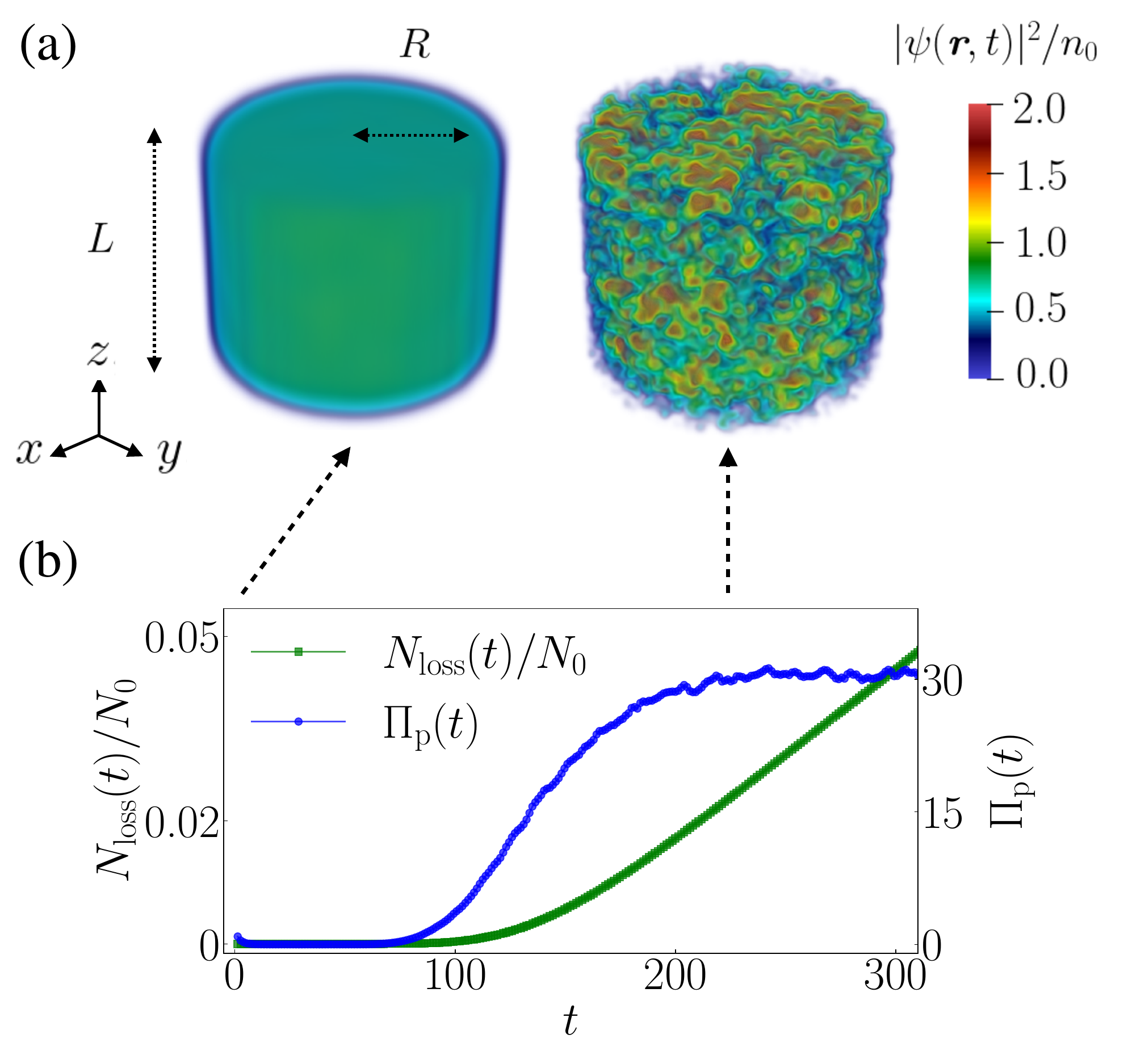}
\caption{Turbulent steady state. (a) Density distributions $|\psi(\boldsymbol r,t)|^2$ for the initial ($t=0$; left) and the turbulent steady ($t=t_\text{ste}$ where $t_\text{ste}\equiv 225$; right) states. Both distributions are normalized to $n_0\equiv N_0/(\pi R^2 L)$. (b) Evolution of the particle loss rate $\Pi_{\rm p}(t)$ and ratio $N_{\rm loss}(t)/N_0$ for $U_\text{s} = 1.36$. 
}
\label{snap}
\end{figure}

The box potential $V_{\rm box}(\bm r)$ is 
\begin{align}
V_{\rm box}(\boldsymbol r)=
\begin{cases}
\displaystyle   0  & \left( \displaystyle \it |z| \le \frac{L}{\rm 2},\it \sqrt{x^{\rm 2}+\it y^{\rm 2}}\le R\right); \\[4 pt]
 \displaystyle  U_{\rm D} &\,\rm\left(otherwise\right),
\end{cases}
\end{align}
where $U_{\rm D}$ is the trap depth, $R$ is the radius, and $L$ is the length of the cylindrical box potential (see Figure~\ref{snap}(a)). 

The external forcing potential $V_{\rm osc } (z, t)$ is 
\begin{equation}
V_{\rm osc } (z, t)= U_\text{s}\sin(\it \omega_{\rm res} t) \frac{z}{L}\,.
\end{equation}

The imaginary potential $iV_{\rm diss}(\bm r)$ is 
\begin{align}
V_{\rm diss}(\boldsymbol r)=
  \begin{cases}
   \displaystyle 0  & \left( \displaystyle \it |z| \le \frac{L+\rm 2 \it \delta}{\rm 2},\it \sqrt{x^{\rm 2}+\it y^{\rm 2}}\le R+\delta \right); \\[4 pt]
   \displaystyle V_{\rm E} & \,\rm\left(otherwise\right).
  \end{cases}
\label{eq:diss}
\end{align}
This potential phenomenologically realizes the dissipation relevant to the experiments mentioned above: it effectively dissipates the wavefunction outside of the box. The parameter $\delta$ is used to avoid dissipating the (small) evanescent-like component of the wavefunction outside (but near the border of) the finite-depth box; a previous study showed that the dynamics is largely independent of the precise value of $\delta$ and $V_{\rm E}$ within a reasonable window~\cite{navon2019synthetic}. The dissipation length scale is $2\pi/k_{\rm D}$ with $k_{\rm D} = \sqrt{U_{\rm D}}$, i.e. loss becomes significant for the particles whose kinetic energy exceeds the trap depth, $k \gtrsim k_{\rm D}$.  

To relate our (dimensionless) simulation scales to the physical ones, the length scale of Eq.~(\ref{eq:GP}) is chosen to be the healing length $\tilde{\xi}=\hbar/\sqrt{2m\tilde{g}\tilde{n}_0}$, where $m$ is the atom's mass, $\tilde{g}=4\pi\hbar^2 a_s/m$ ($a_s$ is the $s$-wave scattering length), and $\tilde{n}_0=N_0/\tilde{V}_{\rm box}$ is the average density where $N_0$ is the initial particle number and $\tilde{V}_{\rm box}$ is the (dimensionful) volume of the cylindrical box. The corresponding time and energy scales are $\hbar/(\tilde{g}\tilde{n}_0)$ and $\tilde{g}\tilde{n}_0$. The dimensionless coupling constant is thus $g=\sqrt{(8\pi)^3 \tilde{n}_0a_s^3}$.

For our simulations, we use typical experimental parameters (see e.g.~\cite{navon2019synthetic}): $g=0.11$, $L=22$, $R=13$, and the wavefunction is normalized to $\int |\psi(\boldsymbol r,0)|^2 \text{d}\bm r=N_0=1.1\times 10^5$. The forcing frequency is set to $\omega_{\rm res} = 0.24$, to resonantly excite the lowest-lying axial excitation - the sound wave of wavelength $2L$, or equivalently, of momentum $k_{\rm F} = \pi/L=0.14$~\cite{garratt2019}. The period of the oscillating potential is $T=2\pi/\omega_{\rm res} \simeq 26$. We set $U_{\rm D}=32$, $V_{\rm E}=-5.0$ and $\delta = 3.0$~\cite{navon2019synthetic}. The grid size is $V_{\rm num} = (L_{\rm num})^3 = 40 \times 40 \times 40$ and the number of grid points is $(N_{\rm grid})^3=128 \times 128 \times 128$.

The numerical simulations are done using the pseudo-spectral method with the fourth-order Runge-Kutta time evolution and a time resolution of $10^{-3}$. The initial state is the ground state in the static dissipationless trap ($U_\text{s} = 0$ and $V_{\rm E} = 0$); it is obtained by imaginary time evolution. We then study the turbulent dynamics by propagating  Eq.~(\ref{eq:GP}) in real time with nonzero $U_\text{s}$ and $V_{\rm E}$.

\section{Momentum distributions for the initial and turbulent steady states}
\label{sec: Emergence}

\begin{figure*}[t]
\includegraphics[width=16.5cm]{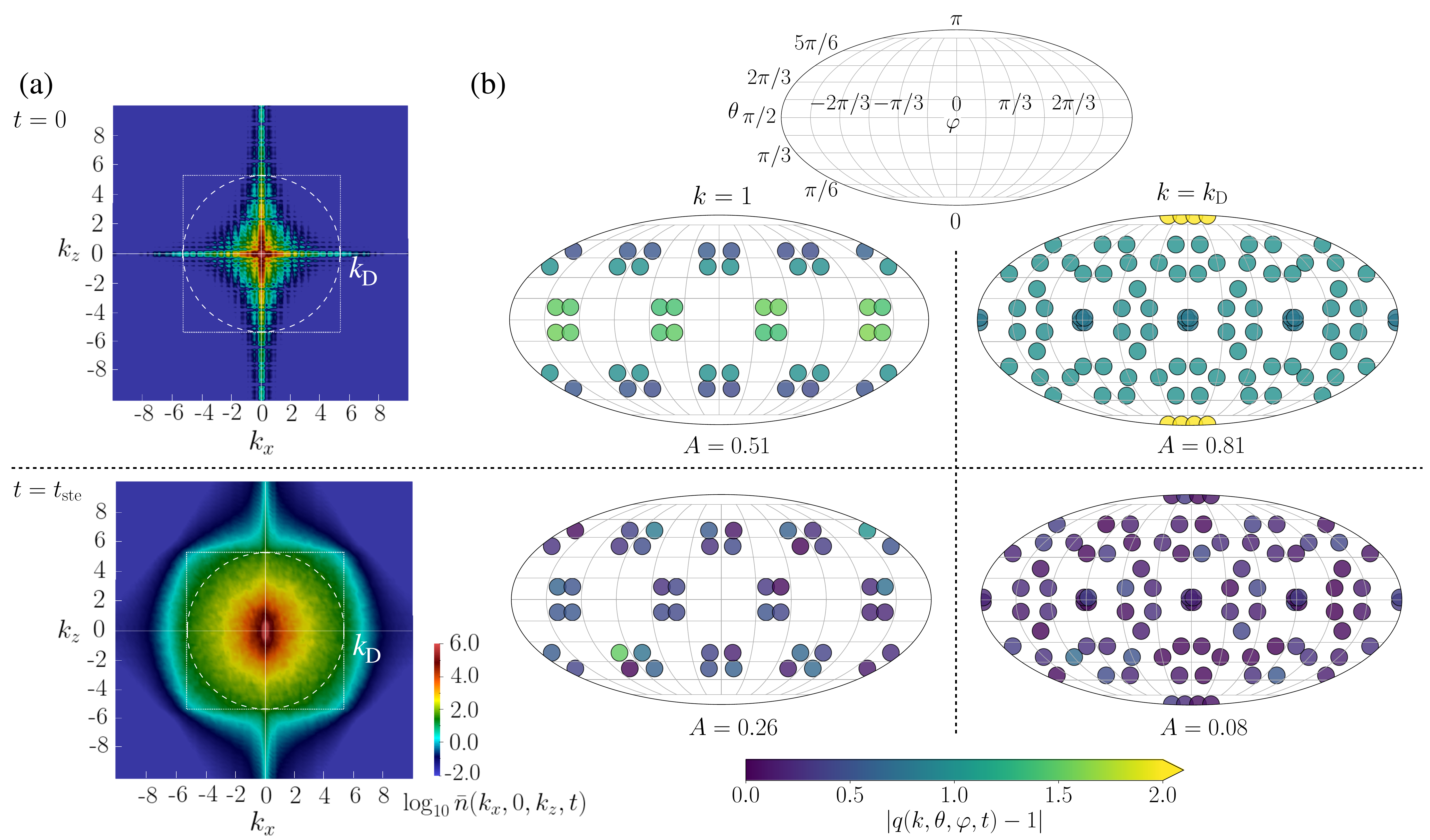}
\caption{Momentum distributions in the initial and turbulent states. (a) Time-averaged momentum distributions $\bar{n}(k_x,0,k_z,t)$ at $t=0$ (the upper panel) and $t_\text{ste}$ (the lower panel). The distributions are plotted on a logarithmic scale. The white dashed (resp. dotted) lines correspond to the condition $|\boldsymbol{k}|=k_\text{D}$ (resp. $\text{min}(k_r,|k_z|)=k_\text{D}$, see text), plotted in the plane $k_y=0$. (b) Mollweide-type projections of the distributions $|q(k, \theta, \varphi, t)-1|$ at $k=1$ and $k=k_{\rm D}$ for $t=0$ (upper panels) and $t=t_\text{ste}$ (lower panels), and the corresponding values of the anisotropy $A(k,t)$. The vertical and horizontal directions are the polar angle $\theta$ and the azimuthal angle $\varphi$ respectively. 
}
\label{mom}
\end{figure*}  

We first perform simulations at $U_\text{s}=1.36$, as shown in Figure~\ref{snap}. The density distribution $|\psi(\boldsymbol r,t)|^2$ is initially quasi-uniform in equilibrium; at long times, it is spatially chaotic (Figure~\ref{snap}(a)). 

We identify the onset of the turbulent steady state by using the particle loss rate $\Pi_{\rm p}(t)$:
\begin{align}
\Pi_{\rm p}(t)\equiv - \frac{\text{d}}{\text{d} t}N(t)\, ,
\label{aloss}
\end{align}
where $N(t)=\int_{\Omega_{\rm box}} |\psi(\boldsymbol r,t)|^2 \text{d}\bm r$ is the total particle number in the box, and $\Omega_{\rm box}= \{ \bm{r} |\   \it |z| \le L/{\rm 2},\it \sqrt{x^{\rm 2}+\it y^{\rm 2}}\le R \}$. Figure~\ref{snap}(b) shows the time evolution of the particle loss $N_{\rm loss}(t)=N_0-N(t)$ and the rate $\Pi_{\rm p}(t)$. At early times, the particle loss and the loss rate are negligible\footnote{A small unimportant particle loss rate seen near $t=0$ is due to a numerical artefact of the switch from imaginary to real-time propagation.}. At $t\approx 100$, $N_{\rm loss}$ starts to rise; correspondingly, the particle-loss rate $\Pi_{\rm p}(t)$ increases. For $t\gtrsim 200$, the loss rate becomes approximately independent of time, and a turbulent steady state is reached \cite{navon2019synthetic}. Note that, strictly speaking, the state is only quasi-steady because the system cannot indefinitely support a constant cascade flux of particles in the presence of dissipation~\cite{navon2016emergence,navon2019synthetic}.

\begin{figure*}[h!]
\centering
\includegraphics[width=16.5cm]{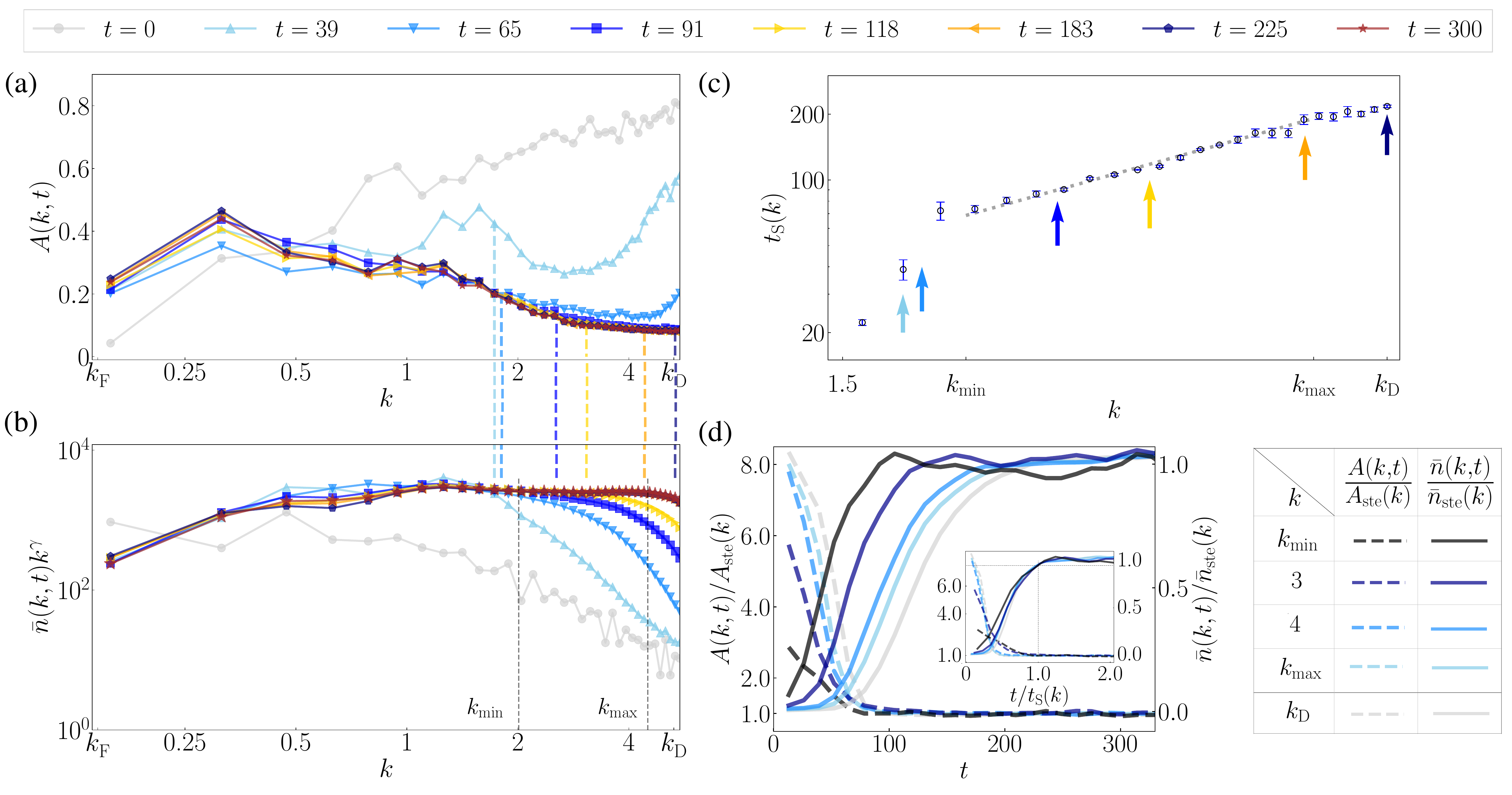}
\caption{Dynamics of the anisotropy and momentum distributions. (a) Time evolution of the anisotropy $A(k,t)$. (b) Evolution of the compensated time-averaged momentum distribution $\bar{n}(k,t)k^\gamma$, with $\gamma=3.75$. The power-law behavior emerges in the inertial range $k_{\rm min} \lesssim k \lesssim k_{\rm max}$ where the anisotropy is small and the exponent is almost constant. (c) The saturation time $t_{\rm S}(k)$ as a function of $k$; the dotted line is a power-law fit. The vertical dashed lines in (a) and (b) correspond to the location of the front $k_{\rm c}$ such that $t_\text{S}(k_{\rm c})=t$. The colored vertical arrows in (c) indicate $k_{\rm c}$ for times corresponding to (a) and (b) (see colors in legend); $k_{\rm c}$ is evaluated by linear interpolation of  $t_{\rm S}(k)$. The error bars on $t_{\rm S}(k)$ are estimated by calculating the saturation time for $\bar{n}_{\rm ste}(k)\pm \delta \bar{n}(k)$, where $\delta \bar{n}(k)$ is the standard deviation of $\bar{n}(k,t)$ over a time $T$ after $t_{\rm S}(k)$.
(d) Time evolution of $A(k,t)/A_{\rm ste}(k)$ and $\bar{n}(k,t)/\bar{n}_{\rm ste}(k)$ for $k=k_{\rm min},3,4,k_{\rm max},$ and $k_{\rm D}$. Inset: same graph with evolution time rescaled to $t_{\rm S}(k)$. 
}
\label{timeave}
\end{figure*}

We next turn our attention to the momentum distribution. It is defined as $n({\bm k},t)= |\Psi(\bm{k},t)|^2$, where $\Psi(\bm{k},t)=(2\pi)^{-3/2}\int \psi(\boldsymbol r,t) \mathrm{e}^{-i\bm k\cdot\bm r} \text{d}\bm r\,$. The grid resolution in $\boldsymbol k$ space is $\Delta k = 2 \pi / L_{\rm num}$ (see Appendix). To avoid that the results depend on the phase of the drive, we compute time-averaged momentum distributions: 
\begin{equation}
\bar{n}(\boldsymbol k,t)=\frac{1}{T}\int_{t-T/2}^{t+T/2}n(\boldsymbol k,t')\text{d}t'\, .
\label{eqtimeave}
\end{equation}
For convenience, we define $\bar{n}(\boldsymbol k,0)\equiv n(\boldsymbol k,0)$. The upper panel of Figure~\ref{mom}(a) shows the initial momentum distribution $\bar{n}(k_x,k_y,k_z,0)$ in the plane $k_y=0$. The distribution is concentrated around the $k_x$ and $k_z$ axes, reflecting the ground state $\psi(\bm{r},0)$ in the box. The momentum distribution of the turbulent state is shown in the lower panel of Figure~\ref{mom}(a), computed at $t=t_\text{ste}$ ($\equiv 225$). The initial sharp features are no longer visible, and as the weight of $\bar{n}$ is larger at high momenta, the distribution becomes more isotropic. A residual anisotropy (along $k_z$) can be seen for $k\lesssim 1$, and is due to the (continuous) anisotropic energy injection along the $z$ axis at $k\approx k_{\rm F}$.

It is interesting to note that even though $\bar{n}(\boldsymbol{k},t_\text{ste})$ is mostly isotropic for $1\lesssim k\lesssim k_\text{D}$, $\bar{n}(\boldsymbol{k},t_\text{ste})$ decays anisotropically for $k> k_\text{D}$ ($k=k_\text{D}$ is shown as dashed white line in Figure~\ref{mom}(a)). This unexpected effect has a geometric origin: in a cylindrical box, a particle with radial momentum $k_r=\sqrt{k_x^2+k_y^2}$ and axial momentum $k_z$ will remain trapped as long as $k_r<k_\text{D}$ and $|k_z|<k_\text{D}$, even though $k$ might be larger than $k_\text{D}$. This condition defines a cylinder in momentum space, whose cut in the $k_r\mathchar`- k_z$ plane is shown as a dotted white line in Figure~\ref{mom}(a). This cut describes well the decaying boundary of $\bar{n}(\boldsymbol{k},t_\text{ste})$ (the small differences might be due to wave- or interaction effects, which are neglected in this simple classical argument).

\section{Measure of anisotropy}
\label{sec: Anisotropydef} We now introduce a measure of momentum-space anisotropy. We define the anisotropy as a distance of the angular distribution at fixed momentum magnitude $k$ to the uniform distribution. Specifically, we first introduce a normalized momentum-dependent angular distribution $q(\boldsymbol{k},t)$:

\begin{equation}
q(\boldsymbol k,t)=\frac{\bar{n}(\boldsymbol k,t)4\pi k^2}{\displaystyle\iint_{S_k} \bar{n}(\boldsymbol k,t)\text{d}S} \, ,
\label{eq:defqk}
\end{equation}
where $\iint_{S_k}\text{d}S=\int_{-\pi}^{\pi}\text{d}\varphi\int_{0}^{\pi}k^2\sin\theta \text{d}\theta$, $(k,\theta,\varphi)$ are the spherical coordinates in momentum space and $S_k$ is the sphere of radius $k$. If the momentum distribution is isotropic on the sphere $S_k$, then $q(\boldsymbol k,t)=1$ for all momentum states on that surface. 

We define the anisotropy $a(k,t)$ as a normalized distance of $q(\boldsymbol{k},t)$ to unity: 
\begin{equation}
a(k,t)=\frac{1}{8\pi k^2}\displaystyle\iint_{S_k} |q(\boldsymbol k,t)-1|\text{d}S.
\label{eq:eqani}
\end{equation}
We have\footnote{$a(k,t)=\iint_{S_k}|q(\boldsymbol{k},t)-1|/(8\pi k^2)\text{d}S < \iint_{S_k}(q(\boldsymbol{k},t)+1)/(8\pi k^2)\text{d}S=1$.} $a(k,t) < 1$, and $a(k,t)=0$ for the isotropic distribution $q(\boldsymbol{k},t) = 1$. To reduce spurious fluctuations on the scale of the discrete momentum grid, we compute a coarse-grained anisotropy: 

\begin{equation}
A(k,t)=\frac{1}{\Delta k}\displaystyle\int_{k-\Delta k/2}^{k+\Delta k/2} a(k',t)\text{d}k'.
\label{eq:cani}
\end{equation}

\begin{figure*}[h!]
\centering 
\includegraphics[width=16.5cm]{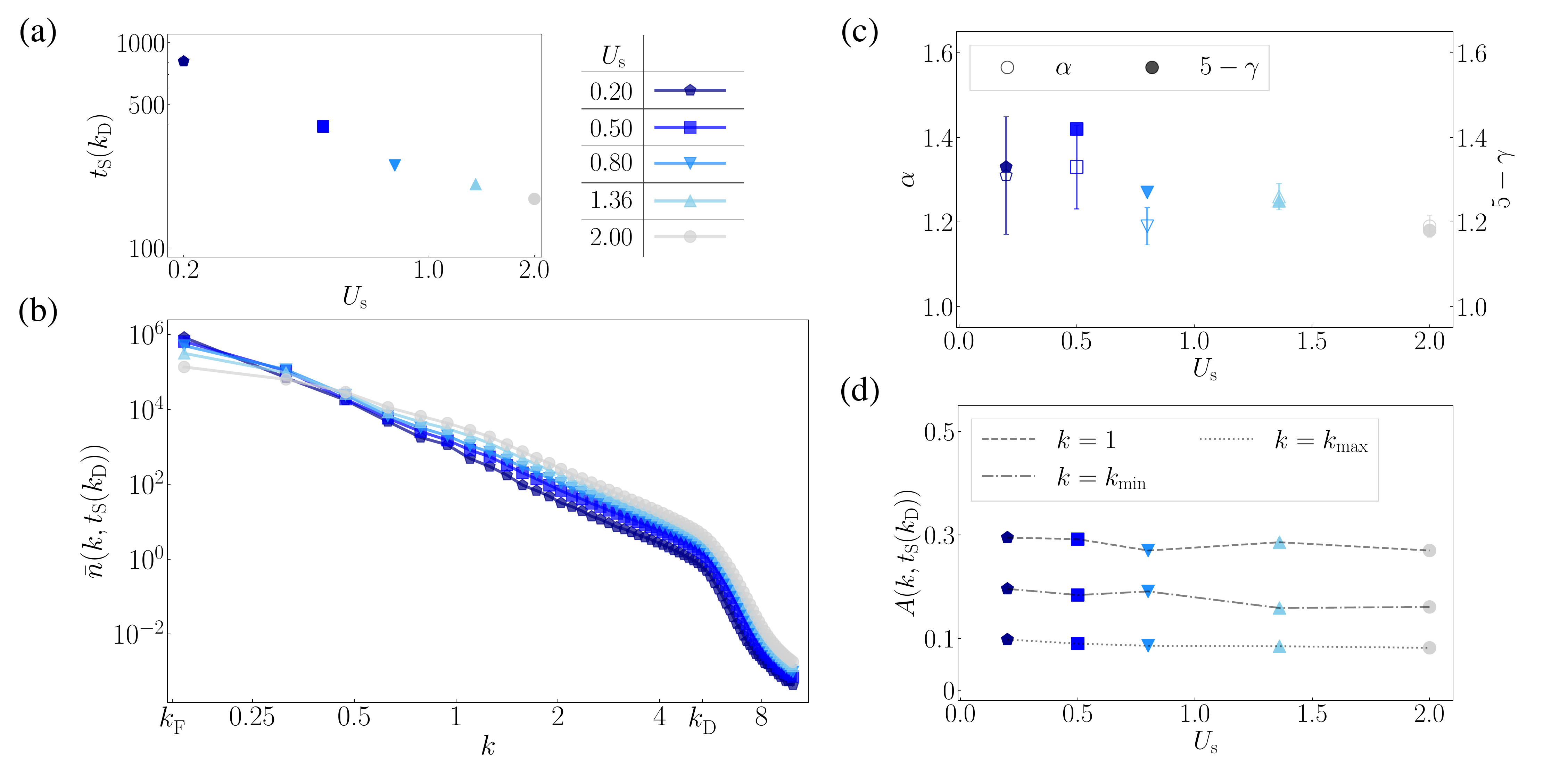}
\caption{Effect of forcing amplitude on the turbulent steady state. (a) Saturation time $t_{\rm S}(k_{\rm D})$ versus $U_\text{s}$. (b) Momentum distributions $\bar{n}(k,t_{\rm S}(k_{\rm D}))$ for each $U_\text{s}$. As the external forcing increases, the occupation numbers for the high momenta become large. (c) The comparison between $\alpha$ and $5-\gamma$ for each $U_\text{s}$. The error bars correspond to the standard error estimated by bootstrapping. (d)  Anisotropy $A(k,t_{\rm S}(k_{\rm D}))$ for the three momenta $k=1$, $k_{\rm min}$, and $k_{\rm max}$. }
\label{steady}
\end{figure*}

In Figure~\ref{mom}(b), we show Mollweide-type projections of $|q(\boldsymbol k,t)-1|$ as a function of the angles $(\theta,\varphi)$ for $k=1$ and $k=k_{\rm D}$ in the initial and turbulent steady state; here, $k=1$ is already quite larger than the forcing momentum $k_{\rm F}$. Initially, $q(\boldsymbol k,0)$ reflects the strongly anisotropic distribution of the wavefunction in the box $\bar{n}(\boldsymbol k,0)$ (see upper panels of Figure~\ref{mom}(a)). This yields high values for the anisotropy: $A(k=1,0)=0.51$ and $A(k=k_{\rm D},0)=0.81$. In the steady state, $q(\boldsymbol{k},t)$ becomes isotropic, as a result of the turbulent dynamics, and $A(k,t_\text{ste})$ decreases to $0.26$ at $k=1$ and $0.08$ at $k=k_{\rm D}$. We attribute a weak (symmetry-breaking) dependence on $\varphi$ to small numerical errors introduced by the chaotic dynamics.

\section{Dynamics of the momentum distribution and the anisotropy towards the turbulent steady state}
\label{sec: anisotropy}

 We now turn to the study of the transient dynamics towards the steady state. In Figure~\ref{timeave}(a), we show the evolution of $A(k,t)$. In the low-momentum region ($k \lesssim 0.3$), the anisotropic forcing dominates, so that the anisotropy over time is always larger than $A(k,0)$.  
 At higher momenta, $A(k,t)$ decreases as turbulence progresses, until it becomes stationary. Unlike $k^{\gamma}\bar{n}(k,t)$, for which the steady state is distinctly developing in the wake of a front propagating in $\boldsymbol k$ space  (see Figure~\ref{timeave}(b) and the vertical dashed color lines), $A(k,t)$ does not seem to evolve in a similar front-like way.
 At long times, $\bar{n}(k,t)$ exhibits a power-law behavior within the inertial range $k_{\rm min}(=2) \lesssim k \lesssim k_{\rm max}(=4.5)$. The exponent in the steady state is $\gamma \approx 3.75$ in that range and is slightly steeper than the prediction for the Kolmogorov-Zakharov (KZ) spectrum of weak wave turbulence~\cite{zakharov2012kolmogorov} (see \href{https://iopscience.iop.org/article/10.1209/0295-5075/aca92e/data}{the supplementary material}).  

To study the front propagation more specifically, we define a momentum-dependent saturation time $t_{\rm S}(k)$ as the earliest time for which $\bar{n}(k,t)$ reaches 95$\%$ of the steady-state\footnote{We define the steady state distribution as $\bar{n}_{\rm ste}(k) \equiv \int_{t_{\rm c}}^{t_{\rm c}+5T}\bar{n}(k,t')/(5T)\text{d}t'$, where $t_{\rm c}$ is a ($U_\text{s}$-dependent) time at which the full momentum distribution has essentially already converged; in practice we use $\int_{k_{\rm min}}^{k_{\rm max}}(\bar{n}(k,t+T/20)-\bar{n}(k,t))^2/\bar{n}(k,t)^2\text{d}k$ of order $10^{-4}$.} $\bar{n}_{\rm ste}(k)$. Figure~\ref{timeave}(c) shows $t_{\rm S}(k)$ as a function of $k$; in the range $k_{\rm min} \lesssim k \lesssim k_{\rm max}$, $t_\text{S}(k)$ scales as a power law of $k$; a fit to the data yields $t_\text{S}(k)\propto k^\alpha$, with $\alpha=1.26\pm 0.03$ (dotted line). In Figure~\ref{timeave}(a)-(b), we indicate for each time series the corresponding momentum for which saturation has occurred, i.e. $k_{\rm c}$ such that $t_\text{S}(k_{\rm c})=t$ (marked with vertical dashed colored lines). Interestingly, the `isotropization' of the momentum distribution precedes the actual cascade front, and its dynamics does not exhibit the self-similar behavior obeyed by the evolution of the momentum distribution. 

We look at this dynamics more closely by plotting the normalized anisotropy\footnote{We define the steady state anisotropy as $A_{\rm ste}(k)\equiv A(k,t_{\rm c})$.} $A(k,t)/A_{\rm ste}(k)$ together with the normalized momentum distribution $\bar{n}(k,t)/\bar{n}_{\rm ste}(k)$ as a function of time in Figure~\ref{timeave}(d), for selected values of $k$. We indeed see that the system becomes isotropic before the momentum distribution reaches its steady-state value. Using the saturation time $t_{\rm S}(k)$ determined previously, we rescale the evolution time as shown in the inset and find that $\bar{n}(k,t/t_{\rm S}(k))/\bar{n}_{\rm ste}(k)$ collapses onto a universal curve, which indicates a self-similar behavior in the inertial range; by contrast,  $A(k,t/t_{\rm S}(k))/A_{\rm ste}(k)$ does not show such self-similarity.

\section{Effect of the forcing amplitude on the turbulent steady state}
\label{sec: robustness}

Finally, we determine the robustness of the steady state isotropy with respect to the forcing amplitude $U_\text{s}$. To compare steady states for various $U_\text{s}$, we first determine the saturation time at $k=k_\text{D}$ as a function of $U_\text{s}$, which we show in Figure~\ref{steady}(a). Secondly, we display in Figure~\ref{steady}(b) the momentum distributions calculated at those saturation times $\bar{n}(k,t_\text{S}(k_\text{D}))$. Aside from an overall factor in the inertial range, the momentum distributions exhibit similar power-law behavior for those forcing amplitudes.

The saturation time $t_{\rm S}(k)$ in the inertial range obeys a power law $t_{\rm S}(k)\propto k^{\alpha}$ (see Figure~\ref{timeave}(c) and Figure~S-2); we show the fitted $\alpha$ versus $U_\text{s}$ in Figure~\ref{steady}(c). Using an argument of energy balance, it was shown in \cite{navon2019synthetic} that for a cascade propagating in momentum space, the exponent $\alpha$ can be related to the exponent $\gamma$ of the momentum distribution; the onset time for losses was shown to scale as a power law of $k_\text{D}$. This argument extends to $k<k_\text{D}$, and we thus expect $t_\text{S}(k) \propto k^{5-\gamma}$. In Figure~\ref{steady}(c), we also show $5-\gamma$, which is in good agreement with the independently-determined $\alpha$. While the prefactor of $t_\text{S}(k)\propto k^\alpha$ depends on $U_\text{s}$, $\alpha$ shows no systematic dependence within our numerical precision.

Finally, we show in Figure~\ref{steady}(d) the anisotropy $A(k,t_{\rm S}(k_{\rm D}))$ as a function of $U_\text{s}$ for three momenta $k=1$, $k_{\rm min}$, and $k_{\rm max}$. Somewhat surprisingly, the anisotropy shows no noticeable dependence on $U_\text{s}$ over a decade, both above and below the bulk chemical potential ($U_\text{s}=1$), further indicating that the steady state is largely insensitive to the details of the drive.

\section{Conclusions}
\label{sec: conclusion}
We studied the emergence of isotropy in matter-wave turbulence using the Gross-Pitaevskii model. We numerically observed how large length scale anisotropy is progressively `forgotten' at smaller length scales as turbulence sets in. In the future, it would be interesting to investigate the linear stability of the KZ solutions of the GP model with respect to anisotropic disturbances~\cite{zakharov2012kolmogorov}, exploiting recent progress on the analytical analysis of such solutions~\cite{Zhu2022direct}. Furthermore, one could extend this work to study more systematically symmetry restoration in the GP model, including spatial homogeneity. Furthermore, ultracold-atom experiments could directly probe anisotropy dynamics, by measuring (either directly or by tomographic reconstruction) the full momentum distributions.

\emph{Note added:} While we completed this manuscript, we became aware of an experimental work studying the emergence of isotropy in a turbulent two-dimensional Bose gas~\cite{galka2022emergence}.

\acknowledgments
We thank T.~Gasenzer, G.~Falkovitch, S.~Nazarenko, L.~Chevillard, and H.~Kobayashi for fruitful discussions. We especially thank K.~Fujimoto for many discussions and comments on the manuscript. Y.~S. acknowledges the support from JST SPRING (Grant No. JPMJFS2138). M.~T. acknowledges the support from JSPS KAKENHI (Grant No. JP20H01855). N.~N. acknowledges support from the NSF (Grant Nos. PHY-1945324 and PHY-2110303), DARPA (Grant No. W911NF2010090), the David and Lucile Packard Foundation, and the Alfred P. Sloan Foundation.

\setcounter{equation}{0}
\renewcommand{\theequation}{A-\arabic{equation}}

\section{Appendix: Computing the anisotropy in discrete-grid momentum space} 
We provide details on the calculation of anisotropy on a discrete numerical grid. 
The grid is a cube of size $(L_{\rm num})^3$ and the real-space coordinate is discretized as $\boldsymbol r_{lmn} = (x_l,y_m,z_n ) = (l-N_{\rm grid}/2,m-N_{\rm grid}/2,n-N_{\rm grid}/2) \Delta x$ with spatial resolution $\Delta x = L_{\rm num}/N_{\rm grid}$ and  grid labels $l,m,n$ taking integers $\{0,\cdots,N_{\rm grid}-1 \}$. Here, $N_{\rm grid}$ is the integer of the grid number in one direction and is assumed to be even in this work. Then, we denote a wavefunction in the real space by $\psi(\boldsymbol r_{lmn})$. Using this notation, we define the discrete Fourier transformation as 
\begin{align}
\begin{split}
\Psi( \boldsymbol k_{\alpha \beta \gamma},t) =& \frac{\Delta x^3}{(2 \pi)^{3/2}} \times\\
&\sum_{l,m,n=0}^{N_{\rm grid}-1} \psi(\boldsymbol r_{lmn},t) 
\mathrm{e}^{ -i \boldsymbol k_{\alpha \beta \gamma} \cdot (\boldsymbol r_{lmn}+L_{\rm num}/2)}.
\end{split}
\end{align}
Here, $\boldsymbol k_{\alpha \beta \gamma} = (\alpha,\beta,\gamma)\Delta k$ is the discrete momentum, with resolution $\Delta k = 2 \pi /L_{\rm num}$; $\alpha,\beta,\gamma$ are integers with values in $\{-N_{\rm grid}/2+1,\cdots, N_{\rm grid}/2\}$. Using $\boldsymbol k_{\alpha \beta \gamma} $, we define the ordered set of discrete (distinct) momenta $\{ |\boldsymbol k_{\alpha \beta \gamma}|\}$ for all allowed values of $\alpha$, $\beta$ and $\gamma$; $k_p$ is defined as the $p$th element of that set (such that $k_0=0$, $k_1=\Delta k$, $k_2=\sqrt{2}\Delta k$, etc.). Then, we numerically evaluate the normalized distribution of Eq.~\eqref{eq:defqk} for $|\Psi(\boldsymbol k_{\alpha \beta \gamma},t)|^2 $ on a sphere of radius $k_p$ using 
\begin{equation}
q(\boldsymbol k_{\alpha \beta \gamma},t)=\frac{|\Psi(\boldsymbol k_{\alpha \beta \gamma},t)|^2 w_1(k_p)}{\displaystyle\sum_{\ (\alpha', \beta', \gamma') \in S_1(k_p)}|\Psi(\boldsymbol k_{\alpha' \beta' \gamma'},t)|^2 },
\end{equation}
where $S_1(k_p) = \{ (\alpha', \beta', \gamma')~|~|\boldsymbol k_{\alpha' \beta' \gamma'}| = k_p  \}$ and $w_1( k_p )=\sum_{(\alpha',\beta',\gamma') \in S_1(k_p)} 1$. 
If $|\Psi(\boldsymbol k_{\alpha \beta \gamma},t)|^2 $ is isotropic on the sphere, $q(\boldsymbol k_{\alpha \beta \gamma},t)$ equals to unity. Then, the anisotropy of Eq.~\eqref{eq:eqani} of the momentum distribution on the sphere of radius $k_p$ is numerically calculated by
\begin{equation}
a(k_p,t)=\frac{1}{2w_1(k_p)}\sum_{(\alpha', \beta', \gamma') \in S_1( k_p)}\left|q(\boldsymbol k_{\alpha' \beta' \gamma'},t)-1 \right|.
\end{equation}
The coarse-grained average of Eq.~\eqref{eq:cani} is calculated by
\begin{equation}
A(k_p,t) = \sum_{p' \in S_2(k_p)} \frac{a(k_{p'},t)}{w_2(k_p)}
\end{equation}
with $S_2(k_p) = \{ p'~|~k_p - \Delta k/2 < k_{p'} \le k_p + \Delta k/2 \}$ and $w_2(k_p)=\sum_{p' \in S_2(k_p)} 1$. Following these formulas, we numerically calculate the anisotropy of the momentum distribution in the main text. 

Note that for the Mollweide-type projections in Figure~\ref{mom}(b),  each grid point has a discrete radial momentum $k_p$, but there are no grid points whose $k_p$ coincides with $k=1$ and $k=k_{\rm D}=5.6195$. Thus, we show the distributions where $k_p$ is $1.0058$ and $5.6199$ for the left and right sides of Figure~\ref{mom}(b), respectively. These momenta are closest to $1$ and $k_{\rm D}$ in our numerical grids.

\end{document}